\documentclass[5p,twocolumn]{elsarticle}

\usepackage{graphicx}
\usepackage{epsfig}
\usepackage{amsmath}
\usepackage{dcolumn}
\usepackage{bm}
\usepackage{dcolumn}
\usepackage{color}
\usepackage{verbatim}
\usepackage{paralist}
\usepackage{amssymb}
\usepackage{url}

\usepackage{lineno}

\makeatletter
\def\ps@pprintTitle{%
 \let\@oddhead\@empty
 \let\@evenhead\@empty
 \def\@oddfoot{}%
 \let\@evenfoot\@oddfoot}
\makeatother

\begin{document}

\begin{frontmatter}
\title{Borexino's search for low-energy neutrino and antineutrino signals correlated with gamma-ray bursts}


\author[GSSI]{M.~Agostini}
\author[Munchen]{K.~Altenm\"{u}ller}
\author[Munchen]{S.~Appel}
\author[Kurchatov]{V.~Atroshchenko}
\author[Milano]{G.~Bellini}
\author[PrincetonChemEng]{J.~Benziger}
\author[Hamburg]{D.~Bick}
\author[LNGS]{G.~Bonfini}
\author[Virginia]{D.~Bravo}
\author[Milano]{B.~Caccianiga}
% ...

\address{\bf{The Borexino Collaboration}}

\address[GSSI]{Gran Sasso Science Institute (INFN), 67100 \L'Aquila, Italy}
\address[Munchen]{Physik-Department and Excellence Cluster Universe, Technische Universit\"at  M\"unchen, 85748 Garching, Germany}
\address[Kurchatov]{National Research Centre Kurchatov Institute, 123182 Moscow, Russia}
\address[Milano]{Dipartimento di Fisica, Universit\`a degli Studi e INFN, 20133 Milano, Italy}
\address[PrincetonChemEng]{Chemical Engineering Department, Princeton University, Princeton, NJ 08544, USA}
\address[Hamburg]{Institut f\"ur Experimentalphysik, Universit\"at, 22761 Hamburg, Germany}
\address[LNGS]{INFN Laboratori Nazionali del Gran Sasso, 67010 Assergi (AQ), Italy}
\address[Virginia]{Physics Department, Virginia Polytechnic Institute and State University, Blacksburg, VA 24061, USA}
% ...

\begin{abstract}
A search for neutrino and antineutrino events correlated with 2,350 gamma-ray
bursts (GRBs) is performed with Borexino data collected between December 2007 and November 2015.
No statistically significant excess over background is observed.
We look for electron antineutrinos ($\bar{\nu}_e$) that inverse beta decay on
protons with energies from 1.8\,MeV to 15\,MeV and set the best limit on the neutrino fluence from GRBs below 8\,MeV. 
The signals from neutrinos and antineutrinos from GRBs that scatter on electrons are also searched for, a detection
channel made possible by the particularly radio-pure scintillator of Borexino.
We obtain currently the best limits on the neutrino fluence of all flavors and species below 7\,MeV. 
Finally, time correlations between GRBs and bursts of events are investigated.
Our analysis combines two semi-independent data acquisition systems for the first time:
the primary Borexino readout optimized for solar neutrino physics up to a few MeV, and a fast waveform
digitizer system tuned for events above 1\,MeV.
\end{abstract}


\end{frontmatter}

\section{Introduction}
\label{sec:intro}
Gamma ray bursts (GRBs) are among the most energetic events known in the Universe, with a typical apparent energy release of $~10^{54}$ erg (or $\sim$1 solar mass), assuming isotropic emission of energy. The average rate of observed GRBs is about 1 event per day from the entire sky. The observer-frame duration of gamma ray emission in the MeV range can be less than 2\,s (for a smaller sub-class of so called short GRBs) or, more typically, is of the order of 10 to 100 seconds. Longer afterglows in X-rays, optical, and radio wavelengths are also observed. The measured redshifts of optical afterglows ($z= 0.01..8.2$) allow to attribute GRBs as extra-galactic events, whose progenitors lie at cosmological distances.

Currently, there is no universally accepted model of GRBs. However, the long GRBs are usually linked to the rotating cores of very massive stars collapsed into neutron stars (NS) or black holes (BH) \cite{Lee 2006, Meszaros 2006}. Short GRBs can result from binary mergers of NS + NS or NS + BH \cite{Meszaros 2006}. These models usually assume that neutrino cooling dominates over the electromagnetic one with neutrino energies in the MeV range \cite{Sahu 2005}. In these models, the energy emitted in the form of MeV thermal neutrinos is of the order of one Solar mass ($M_\odot c^2\approx 2\cdot10^{54}$~ergs) while the energy released in gamma quanta is up to $\sim$100 times less~\cite{Halzen 1996, Meszaros 2015}. Even for the nearest GRBs with $z \sim 0.01$ (for the standard $\Lambda$CDM cosmological model~\cite{Planck}, this comoving distance is $\sim 10^{26}\, {\rm cm}$ or 30 Mpc), the expected fluence of low-energy neutrinos at Earth is equal to $10^5-10^6\, {\rm cm}^{-2}$, which is too small to be observed by current detectors. However, there exist other models for the origin of GRBs, such as cusps of superconducting cosmic strings~\cite{Berezinsky 2001} which can better explain some peculiarities of GRBs~\cite{Cheng 2010}. Some of these models are predicting the fluence of MeV-range neutrinos to be up to $10^{10}$ times larger than that of gamma quanta. The neutrino fluence in these models can be estimated as~\cite{Halzen 1996}:
\begin{equation}\label{eq:eq0}
\begin{split}
\Phi_\nu &= 10^8\,{\rm cm}^{-2}\, \left(\frac{\eta_\gamma}{10^{-10}}\right)^{-1} \left(\frac{E_\nu}{100\,{\rm MeV}}\right)^{-1} \\
& \times\left(\frac{F_\gamma}{10^{-6}\,{\rm erg\cdot cm}^{-2}}\right),
\end{split}
\end{equation}
 where $\eta_\gamma$ is the ratio of photon and neutrino fluences, $F_\gamma$ is the observed gamma ray energy fluence of the GRB. Thus, for a typical GRB with gamma energy release of $3\times 10^{51}\,{\rm erg}$ at redshift $z = 2$ ($F_\gamma=10^{-6}\,{\rm erg\cdot cm}^{-2}$), the predicted fluence of neutrinos with energy of $\sim$10~MeV is $\sim$$10^8 -10^9\, {\rm cm}^{-2}$. For comparison, the {\it observed} flux of solar neutrinos is about $6\cdot 10^{10}\, {\rm cm}^{-2}\cdot{\rm s}^{-1}$, of geo-antineutrinos is about $5\cdot 10^6\, {\rm cm}^{-2}\cdot{\rm s}^{-1}$~\cite{BX2015geonu}. It demonstrates that the sensitivity level of existing neutrino detectors in the MeV range is close to the fluxes expected in several GRB models, if one uses a big set of GRBs. Hereinafter, the energy of neutrino refers to the observed energy; the emitted energy has to be multiplied by factor $1+z$ which is not important due to large uncertainties in models predictions.

Production of TeV and PeV neutrinos by protons accelerated by the plasma shock wave
of GRBs was discussed \cite{Meszaros 2001, Waxman 2007} and such high-energy neutrinos were
searched for by AMANDA~\cite{Achterberg 2007}, ANITA~\cite{Vieregg 2011}, ANTARES~\cite{ANTARES1 2013, ANTARES2 2013}, Baikal~\cite{Avrorin 2009}, IceCube ~\cite{Abbasi 2010, Aartsen 2015}, and SuperKamiokande~\cite{Fukuda 2002}, but no signal was found.
The searches for GRB neutrinos in the MeV energy range have been performed by four
experiments: SuperKamiokande \cite{Fukuda 2002}, SNO \cite{Aharmim 2014}, KamLAND
\cite{Asakura 2015}, and BUST~\cite{BAKSAN2015}.

The SuperKamiokande collaboration searched for electron and muon neutrinos and
antineutrinos in the energy range of 7--80\,MeV. The SNO collaboration
searched for electron neutrinos, electron antineutrinos, and for (anti)neutrinos of non-electron
flavors in the range of 5--13\,MeV. The KamLAND collaboration set upper limits on electron
antineutrino fluence associated with GRBs with known redshift, in the energy ranges of 7.5--100~MeV and (after the Japanese nuclear reactors were switched off in 2011) of 1.8--100\,MeV. The BUST (Baksan Underground Scintillation Telescope) was sensitive to electron neutrinos and antineutrinos in the energy interval of 20--100\,MeV.
None of these four experiments found any correlation between GRBs and neutrino events in their detectors.

In this paper, we present a search for possible correlations between GRBs
and (anti)neutrino events for all neutrino flavours in the Borexino detector.

\begin{figure}[t]
\begin{center}
\includegraphics[width = 8 cm] {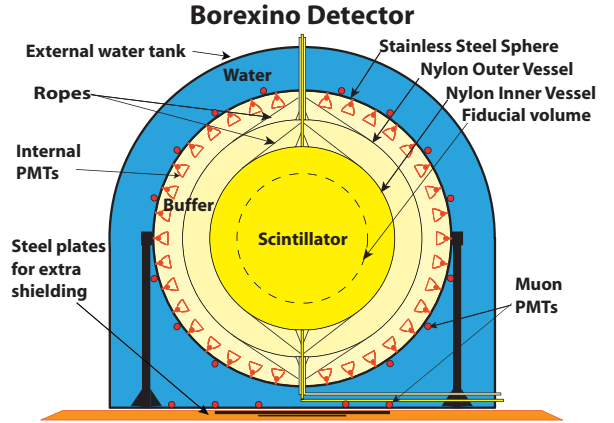}
\caption{Schematic view of the Borexino detector.} \label{fig:Borexino}
\end{center}
\end{figure}

\section{The Borexino detector}
\label{sec:detector}

Borexino is a liquid scintillator detector placed in the Hall C of the underground Laboratory Nazionali del Gran Sasso (LNGS) in central Italy.
The radiopurity of this unsegmented detector has reached unprecedented levels through the techniques described in~\cite{BxFluidHandling}.
This was fundamental in reaching the primary goal of Borexino, the real time spectroscopy of solar neutrinos below 1\,MeV~\cite{LongPaper,Naturepp}.
The detector design~\cite{BXdetector} is based on the principle of graded shielding, with the inner scintillator core at the center of a set of concentric
shells of decreasing radio-purity from inside to outside (see Fig.~\ref{fig:Borexino}).
The active medium, 278\,tons of pseudocumene (PC) doped with 1.5\,g/l of PPO (2,5-diphenyloxazole, a fluorescent dye), is confined within a thin spherical nylon vessel with a radius of 4.25\,m  surrounded by 2212  8" (ETL 9351) photomultipliers (PMTs), defining the so called Inner Detector (ID).
All but 371 PMTs are equipped with aluminum light concentrators designed to increase the light collection efficiency.
The detector core is shielded from external radiation by 890 tons of buffer liquid, a solution of PC and 3-5\,g/l of the light quencher dimethylphthalate (DMP).
The buffer is divided in two volumes by the second nylon vessel with a 5.75\,m radius, preventing inward radon diffusion.
A 6.85\,m radius Stainless Steel Sphere (SSS) encloses the central part of the detector and serves as a support for PMTs.
An external domed water tank of 9\,m radius and 16.9\,m height, filled with ultra--high purity water, serves as a passive shield against neutrons and gamma rays as well as an active muon veto.
The Cherenkov light radiated by muons passing through the water is measured by 208~8" external PMTs defining the so called Outer Detector (OD)~\cite{BxMuons}.

Several calibration campaigns with radioactive sources~\cite{BXcalib} allowed us to decrease the systematic errors of our measurements and to optimize Geant4 based Monte Carlo (MC) simulation.
A typical energy deposit of 1\,MeV produces a signal of about 500 photoelectrons, resulting in an energy resolution of $5\% /  \sqrt {E/\rm{MeV}}$.
A more detailed description of the detector response can be found in~\cite{LongPaper}.

The primary electronics of Borexino, in which all 2212 channels are read individually, is optimized for energies up to few MeV.
For higher energies with a threshold of 1\,MeV, the system of 96 fast waveform digitizers (CAEN v896, 8 bit, thereafter FADC - Flash ADC) was developed, each of them reading-in the signal summed from up to 24 PMTs, with the sampling rate of 400\,MHz.
The details of both systems can be found in~\cite{BXdetector}. In 2009 FADC data acquisition (DAQ) system was upgraded.
Starting from December 2009, it acquires data in a new hardware configuration, having a separate trigger.
FADC energy scale was calibrated using the 2.22 MeV gamma peak originating
from cosmogenic neutron captures on protons, as well as by fitting the
spectrum of cosmogenic $^{12}$B and Michel electrons from muons decaying
inside the detector.
The energy resolution of the FADC system was found to be $10\% /  \sqrt {E/\rm{MeV}}$.
In this work, the first detailed cross-check and a combined analysis based on both systems have been performed.
Advantages of each system, such as higher energy resolution in the primary DAQ or advanced software algorithms for muon and electronics-noise tagging for energies above 1\,MeV in the FADC system, were exploited together.

\section{GRB and time window selection}
\label{sec:data}

\begin{table*}[t]
\caption{Summary of periods and the number of GRBs used in different analyses. The livetimes indicate up-times of different DAQ systems in a given period.}
\begin{center}
\label{table0}

\begin{tabular}{|l|c|c|c|}
\hline
DAQ system & Primary & FADC & Primary + FADC \\
\hline \hline
Period & Dec 2007 - Nov 2015 & Dec 2009 - Nov 2015 & Dec 2009 - Nov 2015 \\
Observed GRBs & 2350 & 1813 & 1813 \\
\hline \hline
Livetime [days] & 2302.0 & 1388.1 & 1279.7 \\

\hline
Used GRBs & 1791 & 1114 & 980 \\
\hline
Analysis & IBD (Sec.~\ref{subsec:IBD}) \& Bursts of events (Sec.~\ref{subsec:bursts}) & IBD (Sec.~\ref{subsec:IBD}) & Elastic scattering (Sec.~\ref{subsec:scatter}) \\
\hline 

\end{tabular}
\end{center}
\end{table*}

We have used the GRB database compiled by the IceCube collaboration~\cite{GRBDB}, which considers the data from several satellites such as {\it SWIFT}, {\it Fermi}, {\it INTEGRAL}, {\it AGILE}, {\it Suzaki}, and {\it Konus/WIND}. This database contains information about GRB's position, time of the detection, duration, energy spectrum, intensity, and, when available, the redshift. We underline that the latter is available only for about 10\% of GRBs. In the period of interest from December 2007 (December 2009) to November 2015, 2350 (1813) GRBs have been observed. 

Considering an upper limit of 0.23\,eV for the sum of neutrino masses~\cite{Planck}, the maximal mass of three neutrino mass states consistent with the measured oscillation mass differences is 87\,meV.
We estimate the time delay, which would be induced in the observation
of gamma rays and 87\,meV neutrinos emitted simultaneously.
For a typical GRB redshift of $z \sim 2$, this time delay reaches 800\,s and then shortly saturates to about 900\,s for higher $z$'s. 
The largest observed GRB redshift is $z$ = 8.2 for GRB 131202A.

Conservatively, we have considered a coincidence time window $\Delta t_{\rm SIG} = \pm 1000$\,s around (before and after) the GRB detection time, in the analysis based on the neutrino-electron elastic scattering channel.
The selected time window covers the possible delay of neutrino signal which propagates slower than photons, as well as the possible earlier emission of neutrinos due to poorly constrained details of GRB physics; the reasonable assumption is that the neutrino burst and the photon burst are not separated by a time much more than the duration of the GRB, which is almost always less than 1000\,s.
We excluded from this analysis those GRBs, for which the fraction of detector's up time covering $\pm 2 \Delta t_{\rm SIG}$ (for reasons discussed below) was less than 95\%.
The rate of events selected as Inverse Beta Decay (IBD) candidates or bursts is much lower with respect to the rate of single point-like events ({\it singles}).
This allowed us, for these two analyses, to extend the $\Delta t_{\rm SIG}$ window to $\pm 5000$\,s (requiring 80\% detector's up time) for a less model dependent search.
In the analyses where the muon veto was applied (Sec.~\ref{subsec:veto}), the $\Delta t_{\rm SIG}$ window in which the candidates (singles, IBDs, bursts) were searched for,  was consequently reduced by $\sim$10\% ($\sim$4300 muons/day passing through the ID). 
In the consideration of the detector's up time coverage of the $\pm (2) \Delta t_{\rm SIG}$ windows, the muon veto was not considered.
The summary of periods and the number of GRBs used in different analyses are given in Tab.~\ref{table0}.

\section{Analysis and results}
\label{sec:results}

We have analyzed Borexino data acquired between December 2007 and November 2015.
The two semi-independent DAQ systems used in this analysis, the primary and the FADC one, have separate triggers and can operate individually, independently from each other.
Different analysis approaches, as it will be specified below, have used data from the two systems in different ways: either individually or in a combined way.
In some cases the presence of the data from both systems was required, in others the FADC system was used in addition to the primary DAQ system only when available.
The total live time of the primary Borexino DAQ system in this period was 2302.0 days.
The FADC system in its new hardware configuration was activated only in December 2009 and was in operation during 1388.1\,days.
Both DAQ systems were working simultaneously for 1279.7 days.

We split the analysis into three independent parts, having in common the definition of muon events (Sec.~\ref{subsec:veto}).
Firstly, we search for an electron antineutrino signal in correlation with GRBs, through IBD reaction with 1.8\,MeV kinematic threshold (Sec.~\ref{subsec:IBD}).
Secondly, we search for all (anti)neutrino flavours, including electron antineutrinos below the IBD threshold: we consider singles, thereby performing neutrino search through elastic scattering off electrons (Sec.~\ref{subsec:scatter}).
Finally, we look for the bursts of events in Borexino data, in correlation with GRBs (Sec.~\ref{subsec:bursts}).

\subsection{Muon veto}
\label{subsec:veto}

In all analyses, muon events have been removed. 
With the exception of the bursts analysis done with a dynamic 2\,s time window (Sec.~\ref{subsec:bursts}), events following every tagged (and removed) muon are excluded from the data sample by applying a 2\,s veto after muons crossing the Inner Detector (ID), and a 2\,ms veto for those crossing only the Outer Detector (OD).
The standard Borexino muon identification~\cite{BxMuons}, including both OD veto and ID pulse-shape muon tagging, is prone to mistakenly tag point-like events above few MeV as muons.
In the previous Borexino analysis (search for antineutrinos from unknown sources~\cite{BxAntinu}) extending up to about 18\,MeV, we have corrected for this effect basing on MC studies.
In this work, we have optimized the muon cuts for higher energies by modifying the muon tagging based on the ID pulse-shape.
In particular, we have employed the anisotropy variable $S_p$~\cite{LongPaper} as well as the reconstructed radius of the event instead of the variables describing the mean duration and the peak position of the light pulse.
In addition, muon tagging based on the FADC waveform analysis was developed and tested against the sample of muons detected by the OD.
The overefficiency to tag point-like events at energies up to few tens of MeV as muons was cross-checked on a sample of Michel electrons.
A slightly decreased muon detection efficiency in the periods when the FADC system was not available is not of concern since in all analysis reported here we search for an excess of events in GRB correlated time windows with respect to GRB uncorrelated times.

\subsection{Inverse beta decay}
\label{subsec:IBD}

In Borexino, electron antineutrinos are detected via IBD reaction $\bar{\nu}_e+p\to n+e^+$ with a threshold of $E_{\bar \nu} = 1.8$\,MeV. The positron thermalizes and annihilates, giving a prompt event. The neutron thermalizes and is captured ($\tau = 254.5 \pm1.8$\,$\mu$s) in most cases by a proton with emission of 2.22\,MeV $\gamma$-ray, defining a delayed event.
In about 1\% of cases, neutrons are captured on $^{12}$C, producing 4.95\,MeV $\gamma$-rays.

The coincidences were searched for with $\Delta t$ = 20 (30) - 1280\,$\mu$s for primary (FADC) systems, excluding muons and muon induced events.
The energy of the prompt event was required to be above the value corresponding to the IBD threshold, considering the energy resolution.
For the delayed event, the energy cut was tuned to cover the gammas from neutron captures both on protons and $^{12}$C.
It was required that no event with energy above 1\,MeV would be present neither 2\,ms before(after) the prompt(delayed) event nor between the prompt and the delayed event.

The position reconstruction is currently available only from the primary Borexino data, while it is not for FADC data.
In order to cross check the consistency between both DAQ systems, we have first performed coincidence searches without  application of position cuts.
 In the final analysis of the primary DAQ data, we have included the requirement of the mutual distance between the prompt and the delayed candidate to be $dR < 1$\,m.
 In total, the live time in the analysis of the primary(FADC) data was 2302.0(1388.1)\,days. 
After excluding GRBs with data coverage of the $\Delta t_{\rm SIG}$ window less than 80\%, we have 1791(1114) GRBs left in the analysis.

Search for correlation with GRBs was performed in time window $\Delta t_{\rm SIG} = \pm5000$\,s around each GRB.
The sum of these time windows defines the integral signal time $T_{\rm SIG}$.
All identified coincidences, considering antineutrinos with energies up to $E_{\bar{\nu}_e} = 15$\,MeV, were divided into 1\,MeV bins according to $E_{\bar{ \nu}_e}  = E_{\rm prompt} + 0.784$\,MeV, where $E_{\rm prompt}$ is the energy of the prompt event.
The procedure described below was then repeated for each energy bin separately.

All antineutrino-like events are divided into those detected within $T_{\rm SIG}$, $N_{\rm in}$ and outside of $T_{\rm SIG}$, $N_{\rm out}$.
All $N_{\rm out}$ candidates are assumed to be accidentals. The background rate $R_{\rm bgr}$ is calculated as:
\begin{equation}\label{eq:eq1}
R_{\rm bgr}=\frac{N_{\rm out}}{T_{\rm total}-T_{\rm SIG}},
\end{equation}
where $T_\textrm{total}$ is the total live time of 2063.3 (1249.9) days for the primary (FADC) DAQ systems. 
Both $T_\textrm{total}$ and $T_\textrm{SIG}$ are after the application of the muon veto cut.

The number of background events $N_{\rm bgr}$ in $T_\textrm{SIG}$ is estimated as: 
\begin{equation}\label{eq:eq1}
N_{\rm bgr}=R_{\rm bgr}T_\textrm{SIG}.
\end{equation}

The 90\% C.L. upper limit $\mu_{90}$ for the number of observed GRB-correlated events is calculated according
to the Feldman-Cousins procedure~\cite{Feldman 1998} with $N_{\rm in}$ and $N_{\rm bgr}$ as input parameters.
The upper limit $N^{IBD}_{90}$ for the number of correlated events for one GRB is then:
\begin{equation}
\label{eq:eq2}
N^{IBD}_{90}=\frac{\mu_{90}}{N_{\rm GRB}},
\end{equation}
where $N_{\rm GRB}$ is the total number of considered GRBs.

We calculate upper limits for the fluence $\Phi_{\bar {\nu}_e}$ for monoenergetic antineutrinos with energy $E_{\bar {\nu}_e}$:
\begin{equation}\label{eq:eq3}
\Phi_{\bar {\nu}_e}(E_{\bar {\nu}_e})=\frac {N^{IBD}_{90}(E_{\bar {\nu}_e}) } {N_p \left < \varepsilon \right > \sigma(E_{\bar {\nu}_e})},
\end{equation}
where $N^{IBD}_{90} (E_{\bar {\nu}_e})$ is the 90\% C.L. upper limit for the number of GRB-correlated events per GRB from the corresponding 1\,MeV energy bin,
$N_p = 1.6 \cdot 10^{31}$ is the number of protons in Borexino scintillator, $\left < \varepsilon \right >$ is the average detection efficiency evaluated via MC simulations for a flat energy spectrum of antineutrinos in the range from 1.8 to 15\,MeV, $\sigma(E_{\bar {\nu}_e})$
is the IBD cross-section for antineutrino energy $E_{\bar {\nu}_e}$ calculated according to~\cite{Strumia 2003}.

\begin{table}[h]
\caption{Borexino 90\% C.L. upper limits for fluence of electron antineutrinos from GRBs.}
\begin{center}
\label{table1}
\begin{tabular}{|c|c|c|}
\hline
$E_{\bar{\nu}_e}$ & $\Phi_{\bar{\nu}_e}$ primary DAQ & $\Phi_{\bar{\nu}_e}$ FADC\\
$\rm{[MeV]}$       & $\rm[{cm^{-2}]}$  & $\rm[{cm^{-2}]}$ \\
\hline \hline
2 & $4.36 \times 10^9$ & $4.87 \times 10^9$  \\
3 & $1.13 \times 10^8$ & $2.64 \times 10^8$   \\
4 & $4.00 \times 10^8$ & $2.49 \times 10^8$   \\
6 & $3.81 \times 10^7$ & $5.67 \times 10^7$   \\
10 & $1.50 \times 10^7$ & $3.57 \times 10^7$  \\
14 & $6.96 \times 10^6$ & $1.04 \times 10^7$  \\
\hline 
\end{tabular}
\end{center}
\end{table}

\begin{figure}[h]
\begin{center}
\includegraphics[width=0.5\textwidth]{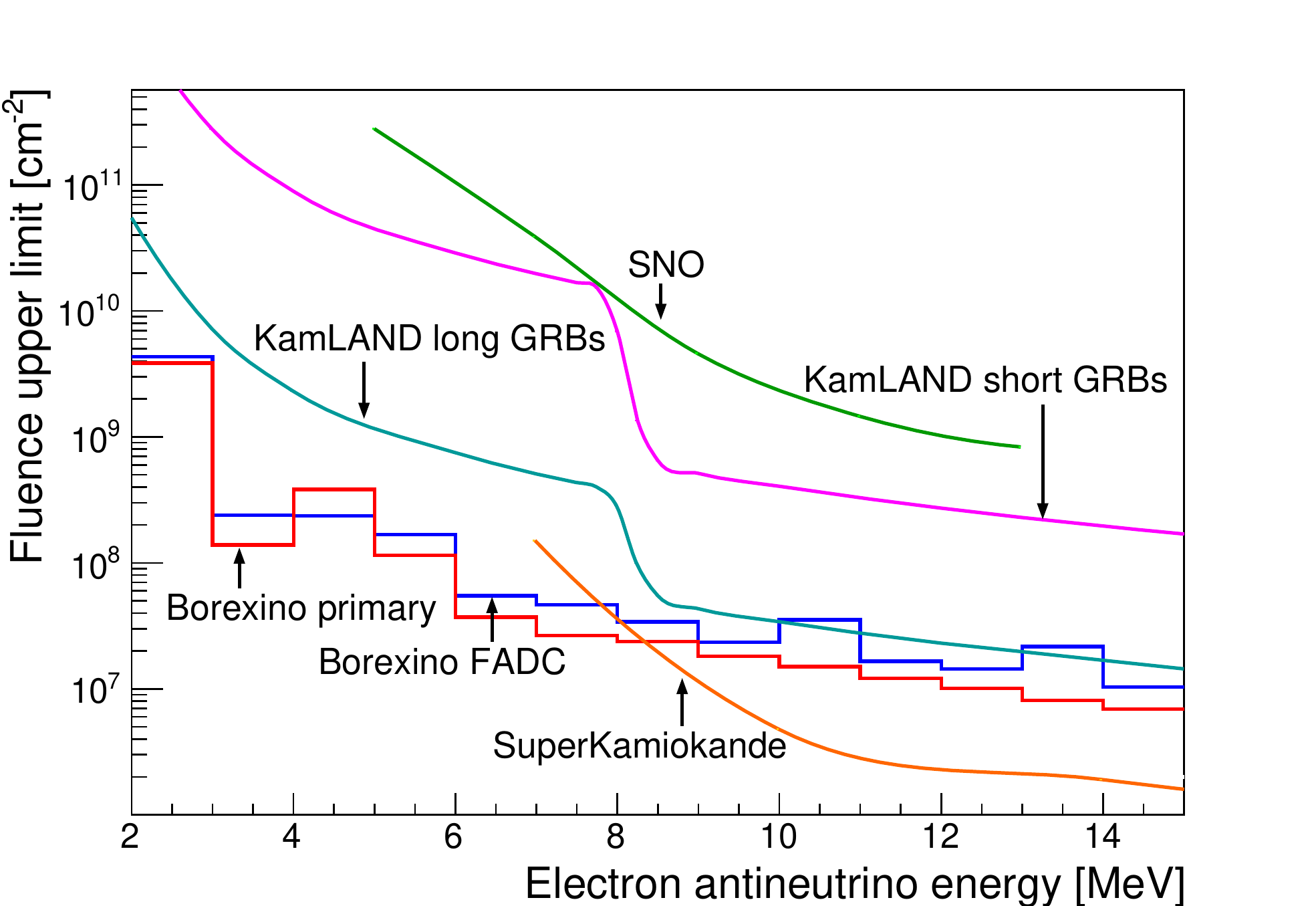}
\caption{Fluence upper limits for electron antineutrinos from GRBs versus antineutrino energy. Borexino results are shown in comparison with results
from SuperKamiokande~\cite{Fukuda 2002}, SNO~\cite{Aharmim 2014}, and KamLAND~\cite{Asakura 2015}.}
\label{cmpr}
\end{center}
\end{figure}

Results for fluence upper limits $\Phi_{\nu}$ for electron antineutrinos, obtained for each energy bin, are given in Table~\ref{table1} and illustrated in Fig.~\ref{cmpr}.
Results previously obtained in other experiments (SuperKamiokande~\cite{Fukuda 2002}, SNO~\cite{Aharmim 2014}, KamLAND~\cite{Asakura 2015}) are also shown in this plot.
The BUST collaboration performed their analysis in the energy range above 20\,MeV~\cite{BAKSAN2015}.
 Borexino limits are the strongest ones in the energy region from 2 to 8\,MeV.

\begin{figure}[t]
\begin{center}
\includegraphics[width = 8 cm] {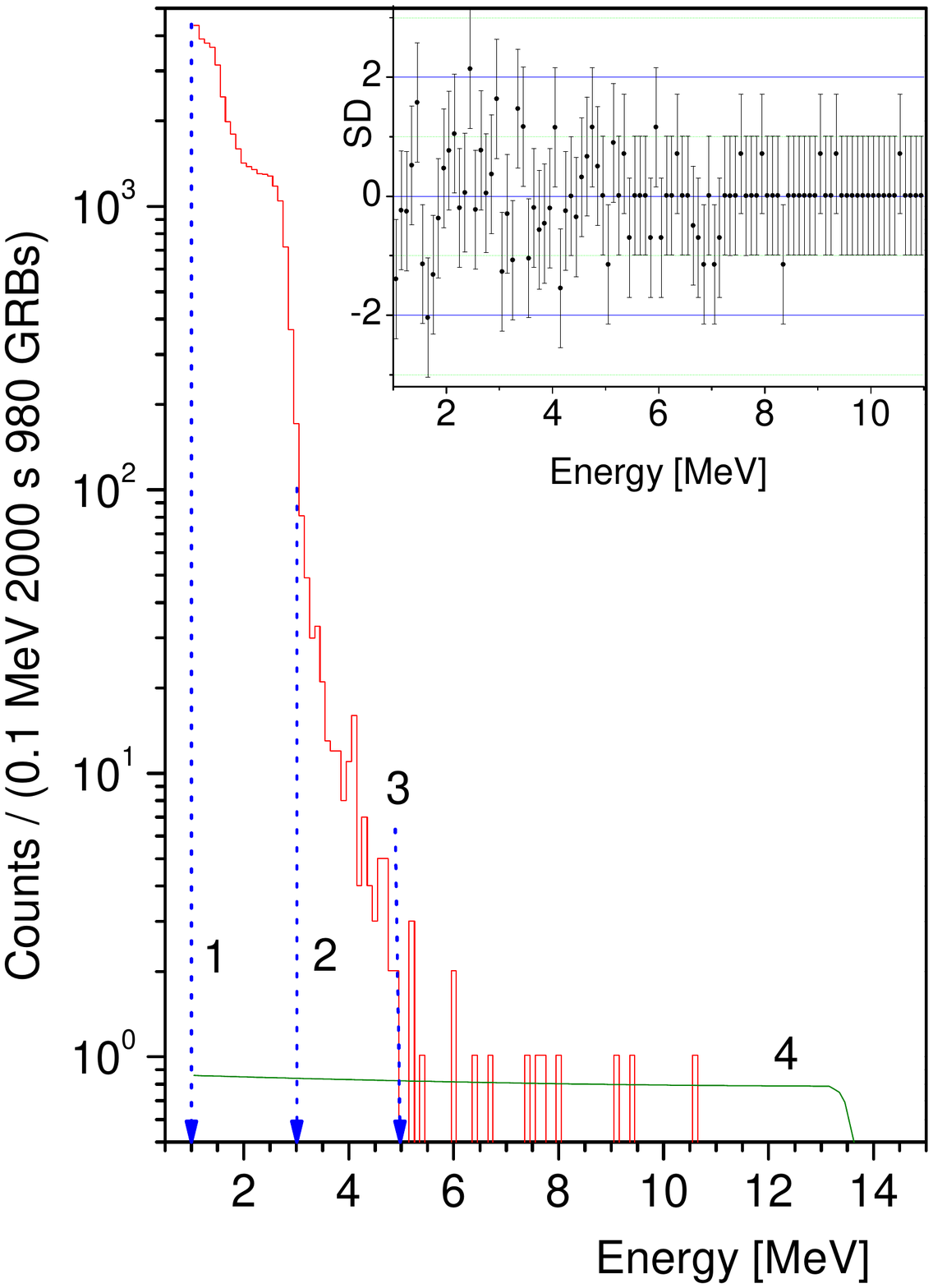}
\caption{Borexino energy spectrum of singles in correlation with GRBs. In inset, the difference between spectra measured in $N_{\rm GRB} \times \Delta t_{\rm SIG}$ and $N_{\rm GRB} \times \Delta t_{\rm BGR}$ time windows is shown in the units of standard deviations (SD). Blue arrows indicate the three energy thresholds $T_{th}$ from Eq.~\ref{eq:ESsigma1} chosen for the separate analysis (details in text). Line 4 shows the expected spectrum of recoil electrons for the fluence $1\times10^{10}~ \rm{cm^{-2}}$ per one GRB of 14\,MeV neutrinos.}
\label{fig:nue_spectrum}
\end{center}
\end{figure}

\begin{figure}[h]
\begin{center}
\includegraphics[width = 8 cm] {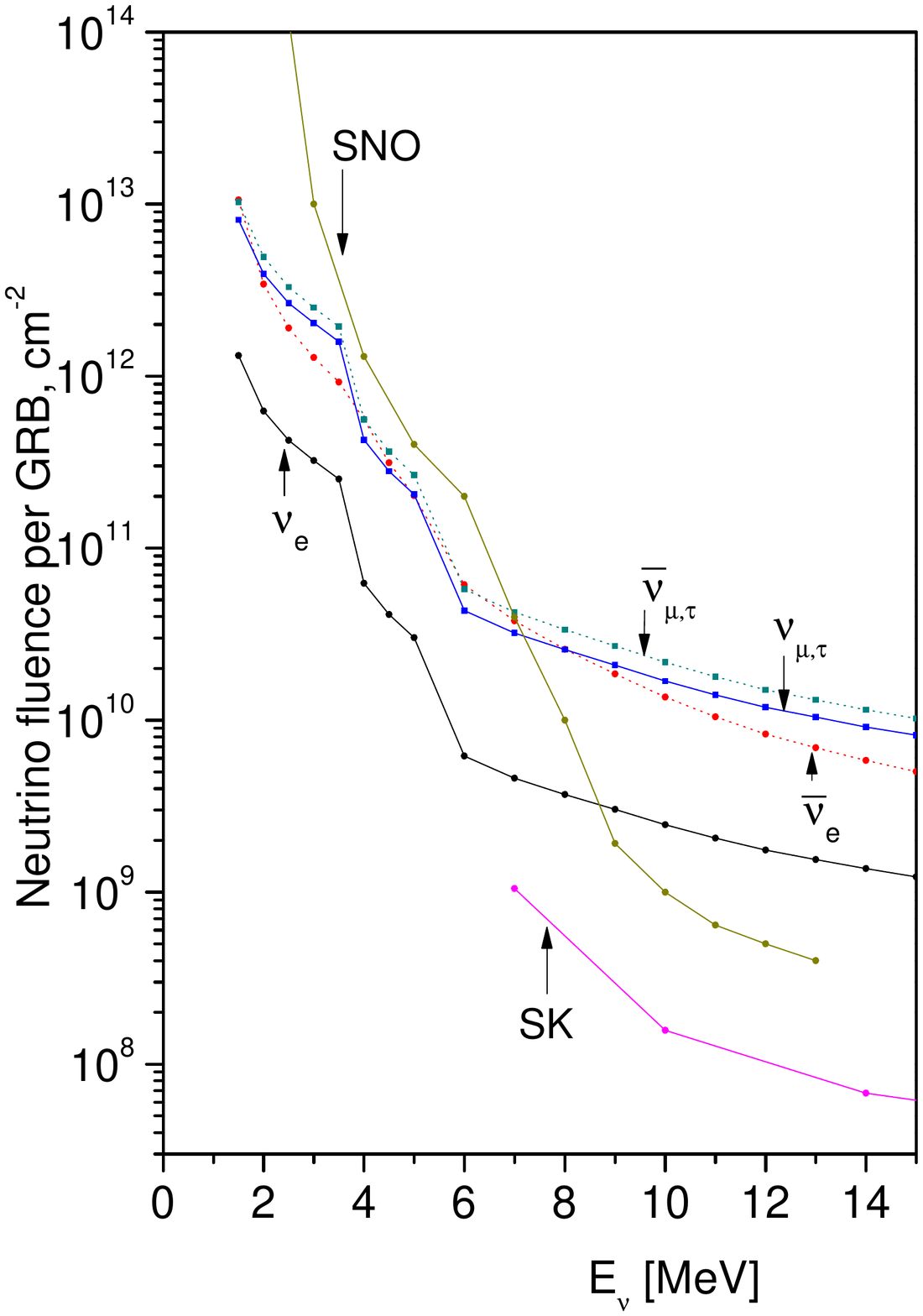}
\caption{Borexino 90\% C.L. fluence upper limits obtained through neutrino-electron elastic scattering for $\nu_{e}$ (line 1), $\bar{\nu}_e$ (2), $\nu_{\mu,\tau}$ (3), and $\bar{\nu}_{\mu,\tau}$ (4). Given are also the limits obtained for $\nu_{e}$ by SNO~\cite{Aharmim 2014} (line 5) and SuperKamiokande~\cite{Fukuda 2002} (line 6).}
\label{fig:nue_limits}
\end{center}
\end{figure}

\subsection{Neutrino-electron elastic scattering}
\label{subsec:scatter}

The goal of this analysis is to search for an excess of singles in Borexino above measured background, in coincidence with GRBs in a time window $\Delta t_{\rm SIG}=\pm1000\;{\rm s}$.
We calculate the overall number of candidate events above 1\,MeV
in $\Delta t_\textrm{SIG}$, excluding muons, muon induced events, and events classified as electronic noise based on FADC pulse shape analysis.

Throughout the whole period of data-taking the counting rate of singles in Borexino is not constant.
In particular, in the energy range above 1\,MeV we observe up to factor of 2 increase in rate during several scintillator purification cycles performed in 2010 -- 2011.
In order to take into account the effect of fluctuations in counting rate, the background was calculated as the number of singles in a 1000\,s time windows, adjacent to $\Delta t_{\rm SIG}$, i.e. $\Delta t_{\rm BGR} = [-2000 .. -1000; 1000 .. 2000]$\,s.
GRBs coverage by Borexino data was then calculated for $\Delta t=\pm2000$\,s window. Requirement of at least 95\% data coverage per window leaves $N_{\rm GRB} = 980$. Overall data coverage in a time window $\pm$2000\,s weighted over 980 remaining GRBs equals to 99.7\%.
This analysis was performed on a data set of 1279.7 days when both the primary and the FADC systems were active.

Figure~\ref{fig:nue_spectrum} shows the energy spectrum of singles measured by the primary DAQ system for the integrated time exposure $N_{\rm GRB} \times \Delta t_\textrm{SIG}$ around the 980 GRBs, in the energy range from 1.5 to 15.0\,MeV. 
The difference between this spectrum and the one measured in the time side bands, $N_{\rm GRB} \times \Delta t_{\rm BGR}$, is shown in the inset. No statistically significant excess is observed in correlations with GRBs.

The scattering of monoenergetic neutrinos with energy $E_{\nu}$ off electrons leads to recoil electrons with a Compton-like continuous energy spectrum with maximum energy $T_{\nu}^{max} = 2 E_{\nu}^2 / (m_e + 2 E_\nu)$.
Thus, the fluence limit for neutrinos of energy $E_{\nu}$ should be calculated by a modified version of Eq.~\ref{eq:eq3}:
\begin{equation}\label{eq:ESfluence}
\Phi_{\nu}(E_{\nu})=\frac {N_{90}^{\nu e} (E_{\nu}) } {N_e \sigma_{\rm {eff}} (E_{\nu})},
\end{equation}
where $N_e = 8.8 \cdot 10^{31}$ is the number of electrons in the Borexino scintillator and $\sigma_{\rm{eff}}(E_{\nu})$ is the effective cross section for neutrinos with energy $E_{\nu}$ via the detection of recoil electrons with reconstructed energy $T$ in the interval ($T_{\rm th} > 0,T_{\nu}^{up})$, with $T_{\nu}^{up} = T_{\nu}^{max} + \sigma_T(T)$ and $\sigma_T$ the energy resolution of the detector, with 100\% detection efficiency and can be expressed as:
\begin{equation}
\label{eq:ESsigma1}
\sigma_{\rm{eff}}(E_{\nu}) = \int _{T_{th}}^{T_{\nu}^{up}} F(T)  {dT},
\end{equation}
where
\begin{equation}
\label{eq:ESsigma2}
F(T) = \int_{T^-}^{T^+}\frac{d\sigma(E_\nu,T')}{dT'} {G(T',\sigma(T');T)}{dT'}.
\end{equation}
The Gaussian function $G(T',\sigma(T');T)$ with variance $\sigma(T')^2$ accounts for the finite energy resolution of the detector, with $T^- = T-3\sigma(T)$ and $T^+ = T+3\sigma(T)$.
The function $F(T)$ is shown as line 4 in Fig.~\ref{fig:nue_spectrum} for $E_{\nu} = 14$\,MeV, $T_{\rm th} = 1$\,MeV, and fluence of $1\times10^{10}~\nu~ \rm{cm^{-2}}$ for one GRB.

The numerator in Eq.~\ref{eq:ESfluence}, $N_{90}^{\nu e} (E_{\nu}) $, is in analogy to Eq.~\ref{eq:eq3}, the 90\% C.L. upper limit on the number of GRB-associated events per one GRB, due to neutrinos with energy $E_{\nu}$, calculated as $N_{90}^{\nu e} = (Q(0.9) \times \sqrt{N_{\rm in} + N_{\rm bgr}}) / N_{\rm GRB}$, where $Q(0.9) = 1.64$ is a quantile function for normal distribution. 
Here, $N_\textrm{in}$ and $N_{\rm bgr}$ denote overall numbers of events in the energy
interval ($T_{th},T_{\nu}^{up}$) detected in the time periods $N_{\rm GRB} \times \Delta t_{\rm SIG}$ and $N_{\rm GRB} \times \Delta t_{\rm
BGR}$, respectively.

The procedure was repeated for neutrino energies $E_{\nu}$ from 1.5 to 5\,MeV in increments of 0.5\,MeV and for $E_{\nu} >$ 5\,MeV in 1.0\,MeV step.
The lower integration limits $T_{th}$ from Eq.~\ref{eq:ESsigma1} was optimized for different neutrino energies considering the shape of the spectrum decreasing with energy (Fig.~\ref{fig:nue_spectrum}).
The three blue lines indicated in Fig.~\ref{fig:nue_spectrum} with indices 1, 2, and 3 show the $T_{\rm th}$, which was set to 1.0, 3.0, and 5.0\,MeV,  for neutrino energies $E_{\nu}$ from 1.5 to 3.5\,MeV, 4.0 to 5.0\,MeV, and 6.0 to 15.0\,MeV, respectively.

In order to set the fluence limits for GRB-correlated neutrinos (antineutrinos) of electron and $(\mu + \tau)$ flavours individually, the corresponding cross section in Eq.~\ref{eq:ESsigma2} was set to $\sigma_{\nu_e}$ ($\sigma_{\bar{\nu}_e}$) and $\sigma_{\nu_{\mu \tau}}$ ($\sigma_{\bar{\nu}_{\mu \tau}}$), respectively, and calculated according to electroweak Standard Model~\cite{SM}.
All limits were obtained in assumption that the whole neutrino flux consists of only individual flavour.
The results obtained for two DAQ systems independently were found to be consistent within statistical uncertainty of our measurements and are summarized in Table~\ref{table2}. 
Figure ~\ref{fig:nue_limits} shows Borexino limits obtained from the primary DAQ ($E_{\nu} <$ 5\,MeV) and the FADC DAQ ($E_{\nu} >$ 5\,MeV). 
Limits for $\nu_{e}$ obtained by SNO \cite{Aharmim 2014} and SuperKamiokande \cite{Fukuda 2002} are shown as well for comparison. 
The Borexino limits are the strongest for all neutrino species below 7\,MeV. 
We note, that for the first time we obtain the limits for electron antineutrinos below IBD reaction threshold through their elastic scattering on electrons.

\begin{table*}[t]
\caption{Borexino 90\% C.L. upper limits for GRB fluences of all neutrino flavours, obtained through the study of neutrino-electron elastic scattering.}
\begin{center}
\label{table2}
\begin{tabular}{|c|c|c|c|c|}
\hline
$E_{\nu}\rm{[MeV]}$ & ${\Phi}_{\nu_e}\rm[{cm^{-2}]}$ & ${\Phi}_{\bar{\nu}_{e}}\rm[{cm^{-2}]}$ & ${\Phi}_{\nu_{\mu,\tau}}\rm[{cm^{-2}]}$ & ${\Phi}_{\bar{\nu}_{\mu,\tau}}\rm[{cm^{-2}]}$\\
\hline \hline
1.5 & $1.31 \times 10^{12}$ & $1.06 \times 10^{13}$ & $8.10 \times 10^{12}$ & $1.03 \times 10^{13}$ \\
2   & $6.25 \times 10^{11}$ & $3.42 \times 10^{12}$ & $3.93 \times 10^{12}$ & $4.93 \times 10^{12}$ \\
3   & $3.23 \times 10^{11}$ & $1.28 \times 10^{12}$ & $2.03 \times 10^{12}$ & $2.50 \times 10^{12}$ \\
4   & $6.24 \times 10^{10}$ & $5.60 \times 10^{11}$ & $4.26 \times 10^{11}$ & $5.60 \times 10^{11}$ \\
6   & $6.18 \times 10^{9}$ & $6.12 \times 10^{10}$ & $4.33 \times 10^{10}$ & $5.77 \times 10^{10}$ \\
10  & $2.46 \times 10^{9}$ & $1.36 \times 10^{10}$ & $1.69 \times 10^{10}$ & $2.17 \times 10^{10}$ \\
14  & $1.37 \times 10^{9}$ & $5.82 \times 10^{9}$ & $9.12 \times 10^{9}$ & $1.15 \times 10^{10}$ \\
\hline
\end{tabular}
\end{center}
\end{table*}

\begin{figure}[ht]
\begin{center}
\includegraphics[width=0.5\textwidth]{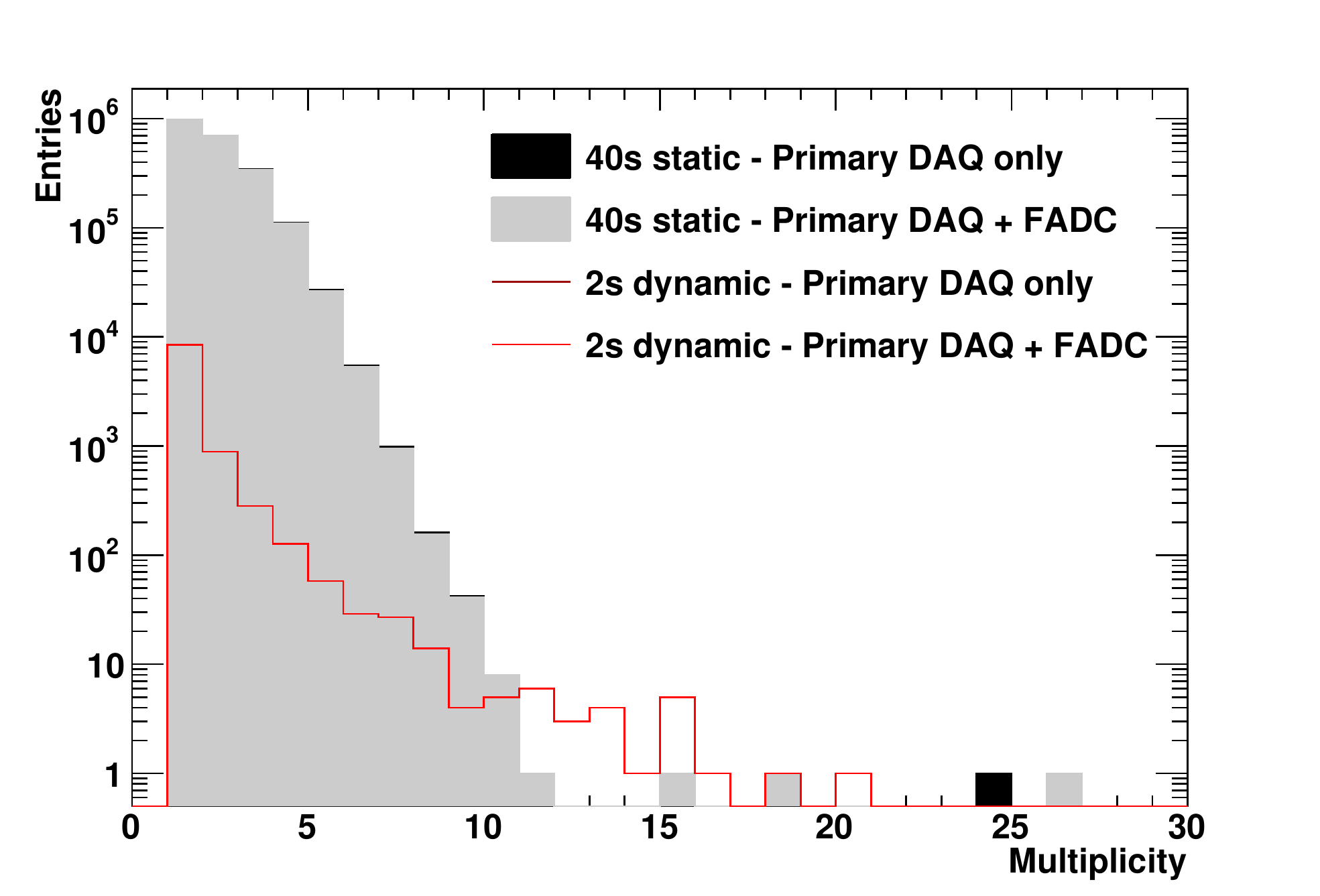}
\caption{The distribution of multiplicities for all bursts of events in Borexino. Note that due to small differences, the results obtained without and with the use of the FADC system cannot be appreciated visually.}
\label{fig:mult}
\end{center}
\end{figure}

\begin{figure}[hh]
\begin{center}
\includegraphics[width=0.5\textwidth]{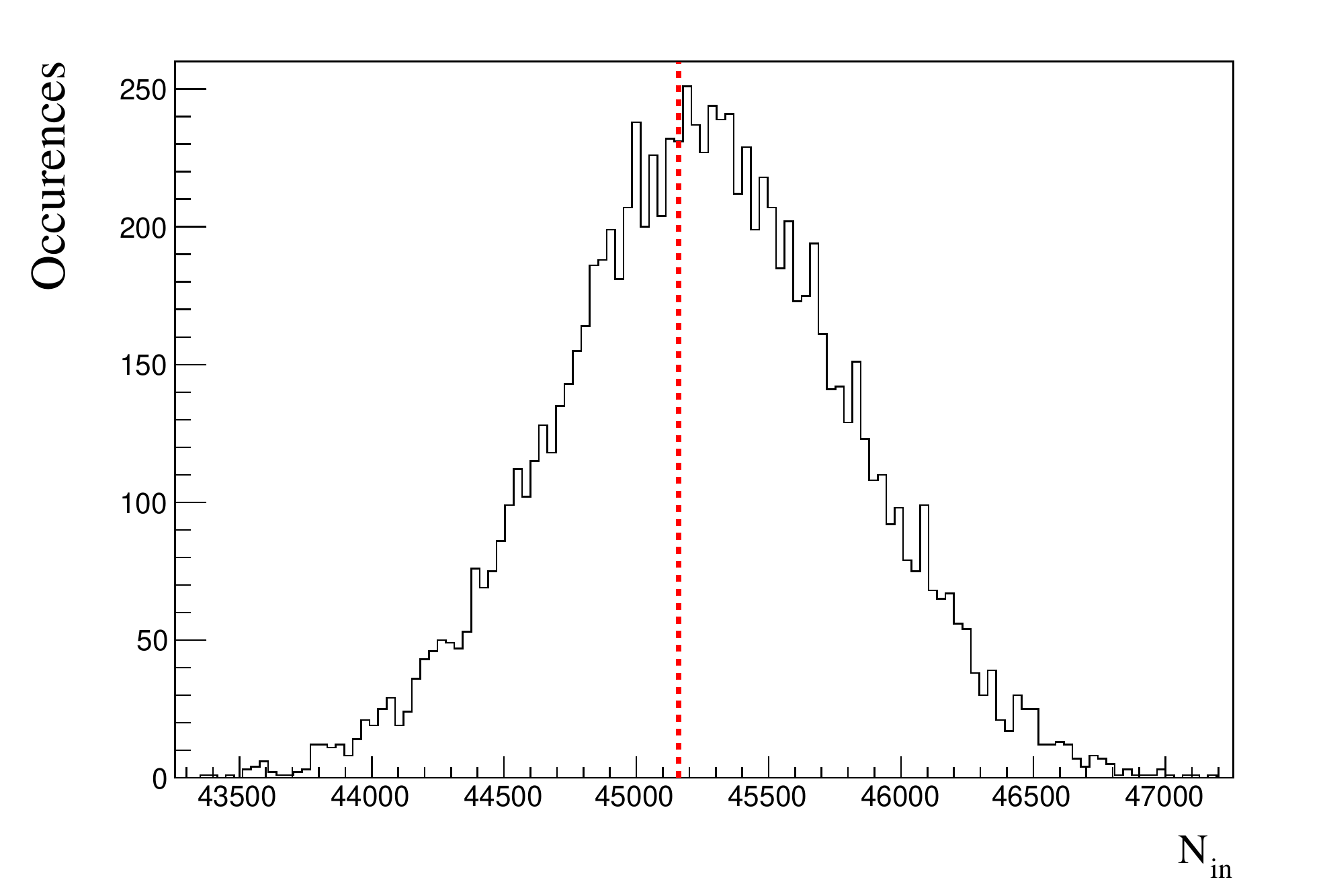}
\caption{Distribution of the number of observed coincidences, $N_{\rm in}$, between the observed bursts with multiplicity 3 (40\,s static time window) and 10000 sets of randomly generated times, equal to the number of GRBs used in this analysis. The vertical dashed red line shows $N_{\rm in}$ corresponding to the real GRB times.}
\label{fig:ToyMC}
\end{center}
\end{figure}

\subsection{Bursts of events}
\label{subsec:bursts}

We have performed a search for bursts of events in Borexino in correlation with GRBs based on 2302.0 days of data from the primary DAQ system. The FADC system was used for additional muon tagging and noise suppression for the time it was available during 1279.7 days.

To define the bursts, we have set two time windows to search for point-like events: a 40\,s static and a 2\,s dynamic.
The 40\,s static window has been chosen regarding the mean duration of GRBs whereas the 2\,s dynamic one covers the maximum duration of short GRBs.

In the static time window strategy, we have searched for non muon-correlated events with energy larger than 1\,MeV. A 40\,s time window is opened after each event and the number of such events which fall within this window is called the burst multiplicity. If one event is present in several windows, it is attributed to the one with the highest multiplicity. If one event is present in several windows with the same multiplicity, it is attributed to the first window.

The main feature of the dynamic time window strategy consists in opening a 2\,s time window after each event with energy larger than 5\,MeV. If another such event falls within this window, another 2\,s time window is opened. This procedure continues until no more event satisfies the selection. Since in Borexino we observe about one muon every 20\,s, in this analysis we did not apply veto after muon events. To suppress cosmogenic background, the energy threshold was increased to 5\,MeV.

In Fig.~\ref{fig:mult} the distributions of multiplicities from both the 40\,s static and the 2\,s dynamic time windows are shown. Considering the bursts with high multiplicities of 8(6), we have detected 10(8) bursts within $\Delta t_{\rm SIG} = \pm$5000\,s time window around GRBs in the static (dynamic) time window approach. Information about bursts of highest multiplicities and in correlation with GRBs for both strategies are shown in Table~\ref{table4}.

To show that all coincidences between bursts of events in Borexino and GRBs are accidental, simple toy MC analysis was performed. Fake GRB times, equal to the number of GRBs used in this analysis, were randomly generated 10000 times for each multiplicity separately. The results show, for both static and dynamic time windows, that the number of observed coincidences between the bursts and GRBs, $N_{\rm in}$, is consistent with the distribution of random coincidences based on the toy MC study. This result is demonstrated for multiplicity~3 in Fig.~\ref{fig:ToyMC}.

\begin{table*}[t]
\caption{Multiplicity, date, time (GMT) and duration for bursts of events in Borexino with the highest multiplicity detected within $\pm$5000s around a GRB. The last column contains names of the GRB which is closest in time. The probability that coincidences of these bursts with GRBs are accidental is 93\% (81\%) for the static (dynamic) time window. }
\begin{center}
\label{table4}
\begin{tabular}{|c|c|c|c|}
\hline
Multiplicity & Date and time (GMT) & Burst's duration, s & Closest GRB\\
\hline \hline
\multicolumn {4} {|l|} {Static 40\,s time window} \\ \hline  \hline
10 & 19 Dec 2010 15:45:00 & 37.1 & 101219B \\
9 &  27 Aug 2014 19:17:41 & 25.8 & 140827A\\\hline
\multicolumn {4} {|l|} {Dynamic time window} \\ \hline  \hline
15 & 19 Dec 2010 15:44:55 & 1.8 & 101219B  \\
12 & 27 Aug 2014 19:17:39 & 1.9 & 140827A\\
\hline
\end{tabular}
\end{center}
\end{table*}

\section{Conclusions}
\label{sec:concl}

We have performed a search for time correlation between gamma ray bursts and events detected by Borexino and associated with neutrinos and antineutrinos reactions -- the inverse beta decay reaction on protons and neutrino-electron elastic scattering. We have also searched for correlations of GRBs with short bursts of events in Borexino. The analysis was performed with data of two semi-independent data acquisition systems: the primary DAQ, optimized for events up to a few MeV, and a Flash ADC system, designed for events above 1\,MeV.

A set of 2350 GRBs observed between December 2007 and November 2015 was checked for correlations with data acquired with the primary DAQ system. A set of 1813 GRBs between December 2009 and November 2015 were also checked for correlations with the data from the FADC DAQ system.
We found no statistically significant time correlations of GRBs with the events in the detector in a time windows of $\pm$5000\,s around GRBs for $\bar{\nu}_e$-like events and for bursts of events, and of $\pm$1000\,s for neutrinos of all species. The demonstrated sensitivity in the MeV range is close to the neutrino fluences predicted by some models of GRBs.
Limits on the fluence of neutrinos of all flavours (and, separately, of electron antineutrinos) were set for neutrino energies 1.5--15\,MeV. 
These are the most stringent bounds for GRB correlated fluence of neutrinos of all species below 7\,MeV, and on the GRB correlated fluence of $\bar{\nu}_e$'s in the range of 2--8\,MeV.

\section*{Acknowledgments}

The Borexino program is made possible by funding from INFN (Italy); the NSF (U.S.);  BMBF, DFG, HGF, and MPI (Germany); RFBR: Grants No. 14-22-03031, No. 15-02-02117, and No. 16-29-13014 (Russia); NCN Poland (Grant No. UMO-2013/10/E/ST2/00180). We acknowledge the generous support and hospitality of the Laboratori Nazionali del Gran Sasso (LNGS).

.

\end{document}